\documentclass{article}
\usepackage{amssymb,amsmath}
\usepackage{epsfig}
\usepackage{subfigure}
\newtheorem{rem}{Remark}[section]
\newcommand{\br}{\begin{rem}}
\newcommand{\er}{\end{rem}}
\newtheorem{ex}{Example}[section]
\newcommand{\bex}{\begin{ex}}
\newcommand{\eex}{\end{ex}}
\newtheorem{Def}{Definition}[section]
\newcommand{\bd}{\begin{Def}}
\newcommand{\ed}{\end{Def}}
\newtheorem{theorem}{Theorem}[section]
\newcommand{\bt}{\begin{theorem}}
\newcommand{\et}{\end{theorem}}
\newtheorem{lemma}{Lemma}[section]
\newcommand{\bl}{\begin{lemma}}
\newcommand{\el}{\end{lemma}}
\newcommand{\be}{\begin{equation}}
\newcommand{\ee}{\end{equation}}
\newcommand{\bea}{\begin{eqnarray}}
\newcommand{\eea}{\end{eqnarray}}
\newcommand{\pa}{\partial}
\newcommand{\nn}{\nonumber}
\newcommand{\adots}{\mathinner{\mkern2mu\raise1pt\hbox{.}\mkern2mu
\raise4pt\hbox{.}\mkern2mu\raise7pt\hbox{.}\mkern1mu}}

\title{First Integrals from Conformal Symmetries:\\  Darboux-Koenigs Metrics and Beyond}
\author{Allan P. Fordy, School of Mathematics, \\
University of Leeds, Leeds LS2 9JT, UK.\\
e-mail a.p.fordy@leeds.ac.uk}

\begin{document}

\maketitle

\begin{abstract}
On spaces of constant curvature, the geodesic equations automatically have higher order integrals, which are just built out of first order integrals, corresponding to the abundance of Killing vectors.  This is no longer true for general conformally flat spaces, but in this case there is a large algebra of {\em conformal} symmetries.  In this paper we use these conformal symmetries to build higher order integrals for the geodesic equations.  We use this approach to give a new derivation of the Darboux-Koenigs metrics, which have only {\em one} Killing vector, but two quadratic integrals.  We also consider the case of possessing one Killing vector and two {\em cubic} integrals.

The approach allows the {\em quantum} analogue to be constructed in a simpler manner.
\end{abstract}



\section{Introduction}

In recent years there has been a burst of activity in the identification and classification of super-integrable systems, both classical and quantum (see the review \cite{13-2} and references therein).
Most of the interest is in Hamiltonians which are in ``natural form'' (the sum of kinetic and potential energies), with the kinetic energy being {\em quadratic} in momenta.  A non-degenerate kinetic energy is associated with (pseudo-)Riemannian metric, so the leading order term in any integral defines a Killing tensor for this metric.  For any Killing tensor, this leading order expression in momenta is itself a first integral for the geodesic equations, since it separately commutes with the kinetic energy.  Thus, before even starting to consider which potentials can be added to a kinetic energy in order for the whole system to be completely (or even super-) integrable, we need to establish which {\em geodesic equations} are themselves completely (or even super-) integrable.

For a manifold with coordinates $(q_1,\dots ,q_n)$, metric coefficients $g_{ij}$, with inverse $g^{ij}$, the geodesic equations are Hamiltonian, with {\em kinetic energy}
\be\label{Ham-h2}
H = \frac{1}{2}\, \sum_{i,j=1}^n g^{ij}p_ip_j,\quad\mbox{where}\quad  p_i=\sum_k g_{ik}\dot q_k.
\ee
For a metric with isometries, the infinitesimal generators (Killing vectors) give rise to first integrals, which are {\em linear} in momenta (Noether constants).
When the space is either flat or constant curvature, it possesses the maximal group
of isometries, which is of dimension $\frac{1}{2}n(n+1)$.  In this case, (\ref{Ham-h2}) is actually the second order {\em Casimir} function of the symmetry algebra (see \cite{74-7}).  Furthermore {\em all} higher order integrals of the geodesic equations are built out of the above Noether constants by just taking polynomial expressions in them.  This is an approach employed in \cite{f07-1,f18-1} for the quantum and classical cases.

Whilst most of the classification results and examples which occur in applications correspond to flat or constant curvature spaces, there are well known examples of conformally flat spaces (but {\em not} constant curvature), possessing quadratic invariants, which are clearly not just quadratic expressions in Noether constants.  Specifically, there are the metrics found by Koenigs \cite{72-5}, which are described and analysed in \cite{02-6,03-11}.
There are other examples of conformally flat spaces (but {\em not} constant curvature), possessing one Noether constant and a {\em cubic} integral (classified in \cite{11-3}), which again {\em cannot} be represented as a cubic expression in the isometry algebra.  The main purpose of this paper is to understand the origin of these integrals in terms of the conformal algebra associated with any conformally flat metric.  Since the isometry group is a subgroup of the conformal group for a given metric and since every constant curvature metric is conformally flat, our approach will include the standard construction of higher order integrals in the constant curvature case.

In 2 dimensions, as is well known, the conformal group is {\em infinite}.  For $n\ge 3$ this group is {\em finite} and has {\em maximal} dimension $\frac{1}{2} (n+1)(n+2)$, which is achieved for {\em conformally flat} spaces
(which includes {\em flat} and {\em constant curvature} spaces).  Any two conformally equivalent metrics have the same conformal group, so we can describe this in terms of the corresponding {\em flat} metric.
In flat spaces, the infinitesimal generators consist of $n$ {\em translations}, $\frac{1}{2} n (n-1)$ {\em rotations}, 1 {\em scaling} and $n$ {\em inversions}, totalling $\frac{1}{2} (n+1)(n+2)$.  This algebra is isomorphic to $\mathbf{so}(n+1,1)$  (see Volume $1$, p143, of \cite{84-4}).  Whilst the conformal algebra in two dimensional spaces is {\em infinite}, there still exists the 6 dimensional subalgebra described above (with $n=2$).  In this paper we only consider 2 dimensional systems, but if this is to be a stepping stone to dealing with higher dimensions, we should only consider this 6 dimensional subalgebra.

In Section \ref{conformal} we list some useful properties of the 6 dimensional conformal algebra of the standard Euclidean metric in 2 dimensions, giving (initially) its Poisson representation in Cartesian coordinates $(x,y)$ and the table of their Poisson relations.  We derive some other coordinate systems, related to commuting pairs, and discuss some 3 dimensional subalgebras, which play a role in our construction.

The main idea of the paper is developed in Section \ref{quadrat-int}, where we consider quadratic {\em conformal invariants} (together with a linear {\em invariant}) and {\em systematically} derive the coefficients corresponding to these being {\em true invariants}.  Inevitably, we are led to the Darboux-Koenigs class of metrics \cite{72-5} (other than flat and constant curvature cases).

In Section \ref{cubic-int}, this idea is used to construct a class of metric with a linear integral and a pair of third order integrals.  We derive Case 3 of the classification given in \cite{11-3}.  Closure of the Poisson algebra leads to an interesting restriction of this metric.

In Section \ref{quantum}, we consider the {\em quantum} extension of the above results.  For the Darboux-Koenigs class of metrics, the quantum and classical formulae are identical at leading order, with some lower order terms being added to some formulae.  However, the quantum version of the system with third order integrals is further restricted to a constant curvature case.  The isometries, as a subalgebra of the conformal symmetries, are identified and expressions for the integrals in terms of these are derived.

\section{The 2D Euclidean Metric and its Conformal Algebra}\label{conformal}

Consider metrics which are conformally related to the standard Euclidean metric in 2 dimensions, with Cartesian coordinates $(x,y)$.  The corresponding kinetic energy (\ref{Ham-h2}) takes the form
\be\label{2d-gen}
H = \psi(x,y) (p_x^2+p_y^2).
\ee
As discussed in the introduction, the conformal algebra of this 2D metric is {\em infinite dimensional}, but we can still write down the $\frac{1}{2} (n+1)(n+2)$ (in this case 6) conformal elements
\bea
&& X_1=p_x,\quad X_2=p_y,\quad X_3=yp_x-x p_y,   \nn  \\[-2mm]
&&    \label{con-alg-xy} \\[-2mm]
&&   X_4 = x p_x+yp_y,\quad X_5 = (x^2-y^2) p_x+2 x y p_y,\quad X_6 = 2 x y p_x+(y^2-x^2) p_y.  \nn
\eea
The elements $\{X_i\}_{i=1}^6$ satisfy the Poisson relations depicted in Table \ref{conalg}.
\begin{table}[h]\centering
\caption{The $6-$dimensional conformal algebra}\label{conalg}\vspace{3mm}
\begin{tabular}{|c||c|c|c|c|c|c|}\hline
   &$X_1$    &$X_2$     &$X_3$   &$X_4$   &$X_5$    &$X_6$    \\[.10cm]\hline\hline
$X_1$    & 0 & 0 & $X_2$ & $-X_1$ & $-2 X_4$ & $-2 X_3$      \\[1mm]\hline
$X_2$    & & 0 & $-X_1$ & $-X_2$ & $2 X_3$ & $-2 X_4$   \\[1mm]\hline
$X_3$    & & & 0 & 0 & $-X_6$ & $X_5$       \\[1mm]\hline
$X_4$    & & & & 0 & $-X_5$ & $-X_6$   \\[1mm]\hline
$X_5$    & & & & & 0 & 0    \\[1mm]\hline
$X_6$    & & & & & & 0        \\[1mm]\hline
\end{tabular}
\end{table}

These conformal elements satisfy
\begin{subequations}
\be\label{HXi}
\{H,X_i\} = \lambda_i H,
\ee
where $H$ is given by (\ref{2d-gen}) and
\bea
&& \lambda_1 = \frac{\psi_x}{\psi},\quad \lambda_2 = \frac{\psi_y}{\psi},\quad \lambda_3 = \frac{y\psi_x-x\psi_y}{\psi},\quad  \lambda_4 = \frac{x\psi_x+y\psi_y-2\psi}{\psi},\nn\\[-1mm]
&&   \label{lambdai}  \\[-1mm]
&&    \lambda_5 = \frac{(x^2-y^2)\psi_x+2 x y\psi_y-4 x\psi}{\psi},\quad   \lambda_6 = \frac{2 x y\psi_x+(y^2-x^2)\psi_y-4 y\psi}{\psi}.  \nn
\eea
\end{subequations}

\subsection{Some Associated Coordinate Systems}\label{uv-coords}

Whenever we have $\{X_i,X_j\}=0$, then there exists a coordinate system $(u,v)$ for which $X_i=p_u,\, X_j=p_v$.  They are defined by
$$
\{u,X_i\}=1,\;\;\; \{u,X_j\}=0 \quad\mbox{and}\quad \{v,X_i\}=0,\;\;\; \{v,X_j\}=1.
$$
In the case of $X_1,\, X_2$, we already have the coordinates $(x,y)$.

\paragraph{Corresponding to $\{X_3,X_4\}=0$,} we have coordinates $(u,v)$, given by $u = \arctan\left(\frac{x}{y}\right),\; v = \frac{1}{2} \log(x^2+y^2)$.
The generating function, $S=e^v (\sin u \, p_x+\cos u\, p_y)$, gives
$$
x = e^v \sin u,\quad y = e^v \cos u, \quad p_x = e^{-v} (p_u \cos u+p_v \sin u),\quad p_y = e^{-v} (p_v \cos u-p_u \sin u).
$$
In these coordinates, we have
\bea
&& X_1=e^{-v} (p_u \cos u+p_v \sin u),\quad X_2=e^{-v} (p_v \cos u-p_u \sin u),\quad X_3=p_u,   \nn  \\[-2mm]
&&    \label{con-alg-X34} \\[-2mm]
&&   X_4 = p_v,\quad X_5 = e^{v} (p_v \sin u-p_u \cos u),\quad X_6 = e^{v} (p_v \cos u+p_u \sin u).  \nn
\eea
We later need the conformal factors $\{H,X_i\} = \lambda_i H$, in these coordinates:
\bea
&& \lambda_1 = \frac{e^{-v}(\cos u\, \psi_u+\sin u (\psi_v+2 \psi))}{\psi},\quad \lambda_2 = \frac{e^{-v}(\cos u\,(\psi_v+2 \psi) -\sin u \,\psi_u)}{\psi},\quad \lambda_3 = \frac{\psi_u}{\psi},  \nn\\[-1mm]
&&   \label{lambdaiuv}  \\[-1mm]
&&    \lambda_4 = \frac{\psi_v}{\psi}, \quad \lambda_5 = \frac{e^{v}(\sin u (\psi_v-2 \psi)-\cos u\, \psi_u)}{\psi},\quad   \lambda_6 = \frac{e^{v}(\sin u \,\psi_u + \cos u\,(\psi_v-2 \psi))}{\psi}.  \nn
\eea

\paragraph{Corresponding to $\{X_5,X_6\}=0$,} we have coordinates $(u,v)$, given by $u = -\,\frac{x}{x^2+y^2},\; v = -\,\frac{y}{x^2+y^2}$.
The generating function, $S=-\,\frac{x}{x^2+y^2}\, p_u -\,\frac{y}{x^2+y^2} \, p_v$, gives
$$
x =  -\,\frac{u}{u^2+v^2},\quad y =  -\,\frac{v}{u^2+v^2}, \quad p_x = (u^2-v^2) p_u  + 2 u v p_v,\quad p_y = 2 u v p_u - (u^2-v^2) p_v.
$$
In these coordinates, we have
\bea
&& X_1= (u^2-v^2) p_u  + 2 u v p_v,\quad X_2= 2 u v p_u + (v^2-u^2) p_v,\quad X_3= vp_u-u p_v,   \nn  \\[-2mm]
&&    \label{con-alg-X56} \\[-2mm]
&&   X_4 = -u p_u -v p_v,\quad X_5 = p_u,\quad X_6 = p_v.  \nn
\eea

\subsection{Involutions and Casimir Functions}\label{inv-cas}

The algebra (\ref{con-alg-xy}) possesses a pair of involutive automorphisms:
$$
\iota_1: (x,y)\mapsto (y,x),\qquad  \iota_2:(x,y)\mapsto \left(\frac{-x}{x^2+y^2},\frac{-y}{x^2+y^2}\right),
$$
under which the elements $X_i$ transform as follows:
$$
\begin{array}{|c||c|c|c|c|c|c|}\hline
& X_1 & X_2 & X_3 & X_4 & X_5 & X_6 \\\hline\hline
\iota_1: & X_2 & X_1 & -X_3  & X_4 & X_6 & X_5 \\\hline
\iota_{2}: & X_5 & X_6 & X_3  & -X_4 & X_1 & X_2 \\ \hline
\end{array}
$$
The action of $\iota_1$ is obvious, while that of $\iota_2$ can be seen by comparing the formulae (\ref{con-alg-xy}) and (\ref{con-alg-X56}).

It can be shown that the abstract algebra depicted in Table \ref{conalg} has 2 quadratic Casimirs:
$$
{\cal C}_1 = X_3^2-X_4^2+X_1X_5+X_2X_6, \quad  {\cal C}_2 = 2 X_3 X_4+X_2 X_5-X_1X_6,
$$
which are multiples of the identity in the matrix (adjoint) representation, but both vanish in this Poisson representation, so are {\em quadratic constraints} on the Poisson algebra.  Under the action of the above involutions we have ${\cal C}_1\mapsto {\cal C}_1$ and ${\cal C}_2\mapsto -{\cal C}_2$.

\subsection{$3-$dimensional Subalgebras}\label{subalg}

This conformal algebra contains several $3-$dimensional subalgebras:
\begin{enumerate}
  \item $\mathbf{g}_{(123)}=\{X_1,X_2,X_3\}$ is the Euclidean algebra with Casimir
  $$
  H_0^{123} =  (X_1^2+X_2^2) =  (p_x^2+p_y^2),
  $$
  corresponding to a flat metric.
  \item $\mathbf{g}_{(563)}=\{X_5,X_6,X_3\}$ is the Euclidean algebra with Casimir
  $$
  H_0^{563} =(X_5^2+X_6^2) = (x^2+y^2)^2\,(p_x^2 + p_y^2),
  $$
  corresponding to a flat metric.
  \item $\mathbf{g}_{(145)}=\{X_1,X_4,X_5\}$ is the algebra $\mathbf{sl}(2)$ with Casimir
  $$
  H_0^{145} = (X_4^2-X_1 X_5) = y^2\,(p_x^2 + p_y^2),
  $$
  corresponding to a metric with constant (non-zero) curvature.
  \item $\mathbf{g}_{(246)}=\{X_2,X_4,X_6\}$ is the algebra $\mathbf{sl}(2)$ with Casimir
  $$
  H_0^{246} =   (X_4^2-X_2 X_6) = x^2\,(p_x^2 + p_y^2),
  $$
  corresponding to a metric with constant (non-zero) curvature.
\end{enumerate}
These are clearly conformally equivalent.  Under the action of $\iota_1$, we have
$$
\mathbf{g}_{(123)}\rightarrow \mathbf{g}_{(123)},\quad \mathbf{g}_{(563)}\rightarrow \mathbf{g}_{(563)},\quad  \mathbf{g}_{(145)}\leftrightarrow \mathbf{g}_{(246)},
$$
whilst under the action of $\iota_2$, we have
$$
\mathbf{g}_{(123)}\leftrightarrow \mathbf{g}_{(563)},\quad \mathbf{g}_{(145)}\rightarrow \mathbf{g}_{(145)},\quad  \mathbf{g}_{(246)}\rightarrow \mathbf{g}_{(246)}.
$$

\section{Geodesic Flows in 2D with Linear and Quadratic Integrals}\label{quadrat-int}

Here we consider the Hamiltonian (\ref{2d-gen}), with general function $\psi(x,y)$.  We see that we cannot expect any of $\lambda_i$, given by (\ref{lambdai}) (or even for any linear combination $\sum_{i=1}^6\alpha_i\lambda_i$), to be {\em zero}.  When such a combination is {\em zero}, we have an {\em invariant} (a {\em Killing vector}), not just a conformal invariant (a {\em conformal Killing vector}).  In this section, we consider Hamiltonians (\ref{2d-gen}) which possess exactly {\em one} such linear invariant.
\br
By a theorem of Darboux and Koenigs, if such a metric possesses at least {\em two} Killing vectors, then it possesses {\em three} and the space has {\em constant curvature} (see \cite{02-6,03-11}).
\er
In fact, we will be more restrictive, and only consider systems for which one of the {\em basis elements} $X_i$ is itself an invariant.  In this case we can use either the coordinates $(x,y)$ or one of the $(u,v)$ systems of Section \ref{uv-coords}.

A {\em quadratic conformal invariant} is any expression of the form
\be\label{gen-quad}
F = \sum_{i,j=1}^6 \beta_{ij} X_i X_j + \sigma(x,y) H,
\ee
where $\beta_{ij}$ is any symmetric matrix of ({\em constant}) coefficients and $\sigma(x,y)$ is an arbitrary function of $(x,y)$ (or, indeed, some appropriate coordinates $(u,v)$), which satisfies
$$
\{F,H\} = (\mu_1(x,y) p_x+\mu_2(x,y) p_y) H,
$$
where $\mu_i(x,y)$ are some functions.

We can ask whether there is a choice of $\beta_{ij}$ and $\sigma(x,y)$ for which $\mu_i(x,y)\equiv 0$, in which case {\em $F$ is a quadratic invariant}.  In fact, we have more structure.  If {\em both} $K=X_i$ (for some $i$) {\em and} $F$ are invariants, then so are $\{F,K\}$, $\{\{F,K\},K\}$, etc.  Simple choices of $\beta_{ij}$ lead to simple equations for $\psi(x,y)$ and $\sigma(x,y)$ and quickly lead to the 4 solutions found by Darboux and Koenigs.  Following \cite{02-6,03-11}, we label these $D_1$ to $D_4$.  To avoid the flat and constant curvature cases, the term $\sigma(x,y) H$ in (\ref{gen-quad}) is essential.  All the Darboux and Koenigs metrics reduce to flat and constant curvature cases when the corresponding parameter vanishes.

Such simple choices can be made by considering the 3 dimensional subalgebras given in Section \ref{subalg}.  We choose a subalgebra that contains our linear invariant $K$.  To avoid unnecessary calculations we also employ the involutions $\iota_1$ and $\iota_2$ of Section \ref{inv-cas}, which means we only need to consider $K$ to be one of $X_2,\, X_3$ and $X_4$, and to be elements of the algebras $\mathbf{g}_{(123)}$ or $\mathbf{g}_{(246)}$ (see the comments at the end of Section \ref{subalg}).

\subsection{The Case when $\{H,X_2\}=0$}\label{X2H=0}

In this case we use the coordinates $(x,y)$, so that $X_2=p_y$, with
$$
H = \varphi(x) (p_x^2+p_y^2) = \varphi(x) (X_1^2+X_2^2).
$$
Note that, for any function $a(y)$, we have $\{a(y),H\} = 2 a'(y) \varphi(x) X_2$.

\medskip
Since $X_2$ belongs to two of our subalgebras, there are two subcases.

\subsubsection{The Algebra $\mathbf{g}_{(123)}$, with $\{H,X_2\}=0$}\label{X2g123}

Noting that $\{X_3,X_2\}=X_1$ and $\{X_1,X_2\}=0$, we define $F_2=X_2X_3+a(y) H$, and then
$$
F_1 = \{F_2,X_2\} = X_2X_1+a'(y) H \quad\mbox{and}\quad \{F_1,X_2\} = a''(y) H
$$
should {\em both} be first integrals {\em if} $F_2$ is.  We therefore have
$$
\{a''(y) H,H\}=2 a'''(y) \varphi(x) X_2 H = 0 \quad\Rightarrow\quad a'''(y) = 0.
$$
The condition
\be\label{f2h-D1}
\{F_2,H\}=0 \quad\Rightarrow\quad (-\lambda_3 +2 a_y \varphi) X_2 H = 0,
\ee
where $\lambda_3=\frac{y \varphi_x}{\varphi}$ in this case (see (\ref{lambdai})).  This is a {\em separable equation} for $\varphi(x)$ and $a(y)$, leading to
$$
\varphi(x) = \frac{1}{\alpha x+\beta}, \quad a(y) = -\frac{\alpha}{4}\, y^2.
$$
We then find
$$
F_1 = \{F_2,X_2\} = X_1 X_2 -\frac{\alpha}{2}\, y H,
$$
and
$$
\{F_1,X_2\} = -\,\frac{\alpha}{2}\,  H,\quad \{F_1,F_2\} = 2 X_2^3+\beta H X_2.
$$
Of course, in 2 degrees of freedom we can only have 3 functionally independent integrals (including $H$), so the above 4 integrals satisfy a polynomial constraint:
$$
F_1^2+\alpha F_2 H+X_2^4 -\beta H X_2^2= 0.
$$
We have obtained, in this way, the first of the Darboux-Koenigs systems $D_1$, which is generally non-constant curvature, but reduces to a flat metric when $\alpha=0$.

\br
The choice of $\sigma(x,y)=a(y)$ guaranteed the factorisation of equation (\ref{f2h-D1}), leading to a single equation for the two functions $\varphi(x)$ and $a(y)$.
\er

\br[The symmetry $x\leftrightarrow y$]
There is an equivalent version of this example, using $X_1$ in place of $X_2$, which can be obtained by just switching $x\leftrightarrow y$, not forgetting that $X_3\mapsto -X_3$ under this transformation.  This is just the involution $\iota_1$.
\er

\subsubsection{The Algebra $\mathbf{g}_{(246)}$, with $\{H,X_2\}=0$}\label{X2g246}

Noting that $\{X_6,X_2\}=2 X_4$ and $\{X_4,X_2\}=X_2$, we define $F_2=X_2X_6+a(y) H$, and then
$$
F_1 = \frac{1}{2}\, \{F_2,X_2\} = X_2X_4+\frac{1}{2}\, a'(y) H \quad\mbox{and}\quad \{F_1,X_2\} = X_2^2+\frac{1}{2}\, a''(y) H,
$$
the latter being a first integral if $a'''(y) = 0$.

The condition
\be\label{f2h-D2}
\{F_2,H\}=0 \quad\Rightarrow\quad (-\lambda_6 +2 a_y \varphi) X_2 H = 0,
\ee
where $\lambda_6=\frac{2 x y \varphi_x-4 y \varphi}{\varphi}$ in this case (see (\ref{lambdai})).  This is a {\em separable equation} for $\varphi(x)$ and $a(y)$, leading to
$$
\varphi(x) = \frac{x^2}{a_2-a_1 x^2},\quad  a(y) = a_1 y^2.
$$
We then find
$$
F_1 = \frac{1}{2}\{F_2,X_2\} = X_2 X_4 + a_1 y H,
$$
and
$$
\{F_1,X_2\} = a_1 H+X_2^2,\quad \{F_1,F_2\} = -2 X_2 (F_2+a_2 H).
$$
These elements satisfy the polynomial constraint
$$
F_1^2= (a_1 H+X_2^2) F_2+a_2 H X_2^2.
$$
We have obtained, in this way, the second of the Darboux-Koenigs systems $D_2$, which reduces to a constant curvature metric when $a_1=0$.

\subsection{The Case when $\{H,X_3\}=0$}\label{X3g123}

In this case we use the coordinates $(u,v)$, associated with $\{X_3,X_4\}=0$ and use the formulae (\ref{con-alg-X34}), so that $X_3=p_u$, so we consider
$$
H = \varphi(v) (p_u^2+p_v^2) = \varphi(v) (X_3^2+X_4^2).
$$
Note that, for any function $a(u)$, we have $\{a(u),H\} = 2 a'(u) \varphi(v) X_3$.

In this case we consider the algebra $\mathbf{g}_{(123)}$, noting that $\{X_1,X_3\}= X_2$ and $\{X_2,X_3\}=-X_1$. We define $F_1=X_1X_3+a(u) H$, and then
$$
F_2 =  \{F_1,X_3\} = X_2X_3+ a'(u) H \quad\mbox{and}\quad \{F_2,X_3\} = -X_1X_3+  a''(u) H=-F_1,
$$
if $a''(u) = -a(u)$.

The condition
\be\label{f1h-D3}
\{F_1,H\}=0 \quad\Rightarrow\quad (-\lambda_1 +2 a_u \varphi) X_3 H = 0,
\ee
where $\lambda_1=\frac{e^{-v} \sin u (\varphi_v +2 \varphi)}{\varphi}$ in this case (see (\ref{lambdaiuv})).  This is a {\em separable equation} for $\varphi(v)$ and $a(u)$, leading to
$$
\varphi(v) = \frac{e^{-v}}{\beta e^v-2\alpha}, \quad a(u) = \alpha \cos u.
$$
We then find
$$
F_2 = \{F_1,X_3\} = X_2 X_3 - \alpha \sin u\, H,
$$
and
$$
\{F_2,X_3\} = -F_1,\quad \{F_1,F_2\} = \beta X_3 H.
$$
These elements satisfy the polynomial constraint
$$
F_1^2 +F_2^2 = \beta H X_3^2+\alpha^2 H^2.
$$
We have obtained, in this way, the third of the Darboux-Koenigs systems $D_3$, which is generally non-constant curvature, but reduces to a flat metric when $\alpha=0$.

\subsection{The Case when $\{H,X_4\}=0$}\label{X4g246}

In this case we use the coordinates $(u,v)$, associated with $\{X_3,X_4\}=0$ and use the formulae (\ref{con-alg-X34}), so that $X_4=p_v$, so we consider
$$
H = \varphi(u) (p_u^2+p_v^2) = \varphi(u) (X_3^2+X_4^2).
$$
Note that, for any function $a(v)$, we have $\{a(v),H\} = 2 a'(v) \varphi(u) X_4$.

In this case we consider the algebra $\mathbf{g}_{(246)}$, noting that $\{X_2,X_4\}= -X_2$ and $\{X_6,X_4\}=X_6$.

We first consider $F_1 = X_2 X_4 + a(v) H$ and the condition
\be\label{f1h-D4}
\{F_1,H\}=0 \quad\Rightarrow\quad (-\lambda_2 +2 a_v \varphi) X_4 H = 0,
\ee
where $\lambda_2=\frac{e^{-v}(2 \varphi \cos u - \sin u \varphi_u )}{\varphi}$ in this case (see (\ref{lambdaiuv})).  This is a {\em separable equation} for $a_v = \alpha e^{-v}$, leading to
$$
-2 \cos u\, \varphi +2 \alpha\varphi^2  + \sin u\,  \varphi_u = 0 \quad\Rightarrow\quad   \varphi(u) = \frac{\sin^2u}{\beta-2\alpha \cos u}.
$$
This has given us $F_1 = X_2 X_4 -\alpha e^{-v} H$, with $\{F_1,X_4\}=-F_1$.

The canonical transformation $(v,p_v)\mapsto(-v,-p_v)$ corresponds to the Lie algebraic involution $(X_2,X_4,X_6)\mapsto (-X_6,-X_4,-X_2)$, as well as $H\mapsto H$, leading to
$$
F_2 =  X_6 X_4 - \alpha e^{v} H.
$$
The remaining Poisson relations are
$$
\{F_1,X_4\} = -F_1,\quad \{F_2,X_4\} = F_2,\quad \{F_1,F_2\} = 2 X_4 (\beta H -2 X_4^2).
$$
These elements satisfy the polynomial constraint
$$
F_1 F_2= \alpha^2 H^2 + X_4^2 (X_4^2-\beta H).
$$
We have obtained, in this way, the fourth of the Darboux-Koenigs systems $D_4$, which reduces to a constant curvature metric when $\alpha=0$.

\section{Geodesic Flows in 2D with Linear and Cubic Integrals}\label{cubic-int}

We again consider a Hamiltonian of the form (\ref{2d-gen}), but now possessing {\em linear} and {\em cubic} integrals.  Such metrics were classified in \cite{11-3}, where it was found that there are three cases, each with a specific form of cubic integral.  Here we use the conformal algebra to build the cubic integral corresponding to their Case 3.

Specifically, we consider the case for which $\{X_2,H\}=0$, so we use Cartesian coordinates $(x,y)$ and consider the Hamiltonian
$$
H = \varphi(x) (p_x^2+p_y^2) = \varphi(x) (X_1^2+X_2^2), \quad\mbox{which satisfies}\quad \{X_2,H\}=0.
$$
We again build a cubic {\em conformal invariant} out of the {\em linear elements} $X_i$ and  {\em quadratic elements} of the form $\sigma(x,y)H$ and restrict the coefficients by asking for it to be a {\em true invariant}.

We mirror the structure found in Sections \ref{X2g123} and \ref{X2g246}, but can no longer restrict to a single 3 dimensional subalgebra.  However, we still use the structure
\be\label{XiXjrels}
\{X_3,X_2\}=X_1,\quad \{X_1,X_2\}=0,\quad \{X_6,X_2\}=2 X_4 \quad\mbox{and}\quad \{X_4,X_2\}=X_2.
\ee
We build $F_1$ and $F_2$, satisfying
\be\label{def-rels}
\{F_2,X_2\}=F_1  \quad\mbox{and}\quad \{F_1,X_2\} = X_2 (c_0 H+c_1 X_2^2),
\ee
where this last term is the general cubic integral involving only $X_2$ and $H$.

\subsection{The Integral $F_1$}

The second of (\ref{def-rels}) is a linear equation for $F_1$, which we can solve:
\be\label{F1-1}
F_1 = c_0 y X_2 H +c_1 X_4 X_2^2 + X_1 (c_2 X_2^2+b(x) H),
\ee
where the first two terms are just a ``particular solution'', with the remaining part being in the {\em kernel} of the linear map $F\mapsto \{F ,X_2\}$.
\br
There are 3 other possible elements of the kernel ({\em excluding} $X_2^3$, which is, itself, a first integral):
$$
X_1^2 (k_1 X_1+k_2 X_2) + \gamma(x) H X_2.
$$
Writing $X_1^2 = \varphi^{-1} H-X_2^2$, the first two terms just redefine $c_2$ and $b(x)$, modify $\gamma(x)$ and add a multiple of the first integral $X_2^3$, so we can remove these two terms.  The calculation of $\{F_2,H\}=0$ then gives $\gamma'(x)=0$, so this is just a constant multiple of $X_2 H$, so can also be removed.  Hence, the formula (\ref{F1-1}) is the general solution of $\{F_1,X_2\} = X_2 (c_0 H+c_1 X_2^2)$.
\er
With $F_1$ defined by (\ref{F1-1}), we find
$$
\{F_1,H\} = (2 c_0 \varphi -c_1 \lambda_4-c_2 \lambda_1-2 b_x \varphi) X_2^2H+(2 b_x-\lambda_1 b) H^2 = 0,
$$
where, in this case,
$$
\lambda_1 = \frac{\varphi_x}{\varphi}, \quad \lambda_4 = \frac{x\varphi_x - 2 \varphi}{\varphi},
$$
leading to
$$
b\, \varphi_x=2\, \varphi\, b_x \quad\mbox{and}\quad  (c_2+c_1 x)\varphi_x = 2(c_1+c_0\varphi-b_x\, \varphi)\varphi,
$$
giving
\be\label{bfi-sol}
\varphi = b^2 \quad\mbox{and}\quad   b_x = \frac{b(c_1+c_0 b^2)}{c_2+c_1 x+b^3}\, .
\ee

\subsection{The Integral $F_2$}

Given this solution for $F_1$, we can solve $\{F_2,X_2\}=F_1$ for $F_2$ (using the relations (\ref{XiXjrels})):
\be\label{F2-1}
F_2 = \frac{1}{2}\, c_0 y^2 X_2 H +\frac{1}{2}\,c_1 X_6 X_2^2 + X_3 (c_2 X_2^2+b(x) H)+b_1(x) H X_2,
\ee
where the last term commutes with $X_2$.  We then find
$$
\{F_2,H\} = \left(2 c_0 y\varphi -\frac{1}{2}\, c_1 \lambda_6-c_2 \lambda_3\right) X_2^2H-\lambda_3 b H^2+2 b_x\varphi X_1X_2 H+2 b_{1x}\varphi X_1X_2 H = 0,
$$
where, in this case,
$$
\lambda_3 = \frac{y\varphi_x}{\varphi}, \quad \lambda_6 = \frac{2 y (x\varphi_x - 2 \varphi)}{\varphi}.
$$
If we write
$$
X_3 = y X_1-x X_2, \quad \varphi X_1^2 = H - \varphi X_2^2,
$$
then
$$
\left(2 c_0 \varphi -c_1 \left(\frac{x\varphi_x-2 \varphi}{\varphi}\right)-c_2 \frac{\varphi_x}{\varphi}-2 b_x \varphi\right)\, y X_2^2H+\left(2 b_x-\frac{b\varphi_x}{\varphi}\right) H^2 +2 (b_{1x}-xb_x)\varphi X_1X_2H = 0.
$$
These coefficients give
$$
b_{1x}=x b_x,\quad b\, \varphi_x=2\, \varphi\, b_x \quad\mbox{and}\quad  (c_2+c_1 x+b\, \varphi)\varphi_x = 2(c_1+c_0\varphi)\varphi,
$$
giving, again, (\ref{bfi-sol}) and an additional function $b_1(x)$, obtained by integrating $b_{1x}=x b_x$.

\br
The {\em scalar curvature} of this metric is given by $R=2(b b_{xx}-b_x^2)$.
\er

\br[Equivalence with Case 3 of \cite{11-3}]
In \cite{11-3}, the metric coefficient $\varphi(x)$ is denoted $h_x^2$ and for case (iii) of equation (1.2), $h(x)$ satisfies
$$
h_x (A_0h_x^2-A_1h+A_2) = A_3 x+A_4.
$$
Differentiating this and eliminating $h(x)$ itself, we find
$$
h_{xx} = \frac{h_x (A_3+A_1 h_x^2)}{A_4+A_3 x +2 A_0 h_x^3},
$$
which is exactly our equation (\ref{bfi-sol}) for $b=h_x$, with $c_0=\frac{1}{2}A_1,\, c_1=\frac{1}{2}A_3,\,c_2=\frac{1}{2}A_4$ and $A_0=1$ (absorbed into the definition of $b$).
\er

\subsection{Closing the Poisson Algebra}

We have the relations (\ref{def-rels}), subject to (\ref{bfi-sol}) and $b_{1x}=x b_x$.  The Jacobi identity implies
$$
\{\{F_1,F_2\},X_2\} = \{\{F_1,X_2\},F_2\}=-(c_0H+3 c_1 X_2^2)F_1 = -\{(c_0H+3 c_1 X_2^2)F_2,X_2\},
$$
so we can write
\be\label{f1f2-cub}
\{F_1,F_2\}+(c_0H+3 c_1 X_2^2)F_2 = d_1 H^2 X_2+d_2 H X_2^3+d_3 X_2^5.
\ee
Comparing coefficients of $p_x^i p_y^{5-i}$ in this equation, we find $d_1$ and $d_2$ are arbitrary, $d_3=3 c_2^2$ and
\be\label{b1-form}
b_1 = \frac{2 c_2+2 c_1 x+d_1 b +2 c_0 x b^2 - b^3}{2 c_0 b},
\ee
and $b(x)$ is constrained to satisfy the quartic equation:  {\small
\be\label{b-quartic}
P_4 = 2c_1 b^4-4 c_0(c_2+c_1 x)b^3+(2c_0 x(2 c_2+c_1 x)+2 c_1d_1-c_0d_2)b^2-4 c_1(c_2+c_1 x) b+2 c_0(c_2+c_1x)^2 = 0.
\ee  }
Remarkably, these are consistent with the formulae (\ref{bfi-sol}) and $b_{1x}=x b_x$.  Differentiating (\ref{b1-form}) with respect to $x$ and replacing $b'(x)$ by the formula in (\ref{bfi-sol}), gives
$$
b_{1x} = \frac{xb(c_1+c_0 b^2)}{c_2+c_1 x+b^3},
$$
which is just $b_{1x}=x b_x$.  Differentiating (\ref{b-quartic}) with respect to $x$ and replacing $b'(x)$ by the formula in (\ref{bfi-sol}), gives
$$
P_{4x} = \frac{2 (c_1+c_0 b^2)}{c_2+c_1 x+b^3} \, P_4 = \frac{2 b'}{b}\, P_4,
$$
which allows the solution $P_4=0$.

These integrals satisfy the constraint
\be\label{con-cub}
F_1^2-2 X_2 (c_0H+c_1 X_2^2) F_2 = H^3-c_2^2 X_2^6 -d_1 X_2^2 H^2 -\frac{1}{2}\, d_2 H X_2^4.
\ee

\section{The Quantum Case}\label{quantum}

For a given metric $g_{ij}$, we have the classical kinetic energy and its quantum analogue (the Laplace-Beltrami operator):
$$
H = \frac{1}{2} \, \sum_{i=0}^n g^{ij} p_ip_j \quad \mbox{and}\quad
   L_b f = \sum_{i,j=1}^n g^{ij} \nabla_i\nabla_j f = \sum_{i,j=1}^n \frac{1}{\sqrt{g}} \frac{\pa}{\pa x^j}\left(
               \sqrt{g} g^{ij}\frac{\pa f}{\pa x^i}\right),
$$
where $g$ is the determinant of the matrix $g_{ij}$.  However, there is no such general algorithm for finding the quantum analogue of other quadratic and higher order integrals.  Nevertheless, there \underline{is} an algorithm for finding the quantum analogue of a {\em first order} integral or conformal invariant, and this will be exploited in this section.

For a metric with isometries, the
infinitesimal generators (Killing vectors) are just first order differential
operators which commute with the Laplace-Beltrami operator $L_b$.  When
the space is either flat or constant curvature, it possesses the maximal group
of isometries, which is of dimension $\frac{1}{2}n(n+1)$.  In this case, $L_b$
is proportional to the second order {\em Casimir} function of the symmetry algebra
(see \cite{74-7}).

Recall that if $f,\, g$ are any functions of $({\bf q},{\bf p})$, then the Hamiltonian vector field of $f$ is
\be\label{hamvec}
X_f=\sum_{i=1}^3 \left(\{q_i,f\}\pa_{q_i}+\{p_i,f\}\pa_{p_i}\right) \quad\mbox{and}\quad  [X_f,X_g]=-X_{\{f,g\}}.
\ee
Functions which are \underline{linear} in momenta define vector fields on configuration space, with coordinates $(q_1,\dots,q_n)$.  For any function on configuration space, $f(q_1,\dots,q_n)$, we have:
\be\label{L-def}
h({\bf q},{\bf p}) = \sum_{i=1}^n a_i({\bf q})p_i \quad\Rightarrow\quad L_h f = \{f,h\} = \sum_{i=1}^n a_i({\bf q}) \frac{\pa f}{\pa q_i},
\ee
and it follows from (\ref{hamvec}) that $[L_f,L_g]=-L_{\{f,g\}}$.
This is one of the advantages of expressing integrals in terms of conformal (or even ``true'') invariants.  We can immediately {\em quantise} any classical system which is built in this way.  This idea is used in \cite{f07-1,f17-6}.

\subsection{The Conformal Algebra}

We use (\ref{L-def}) to define $L_i = L_{X_i}$ for each of the elements of the conformal algebra (\ref{con-alg-xy}).  We need expressions in both Cartesian coordinates $(x,y)$ and the coordinates $(u,v)$, related to $\{X_3,X_4\}=0$.

In Cartesian coordinates $(x,y)$, we have
\begin{subequations}
\bea
&&  L_1=\pa_x,\quad L_2 = \pa_y,\quad L_3 = y \pa_x-x \pa_y,\quad L_4 = x \pa_x+y \pa_y,\nn\\[-1mm]
&&   \label{con-algL-xy}  \\[-1mm]
&&  L_5 = (x^2-y^2)\pa_x+2 x y \pa_y, \quad L_6 = 2 x y \pa_x + (y^2-x^2) \pa_y.   \nn
\eea
These satisfy the commutation rules $[L_i,L_j]=-L_{\{X_i,X_j\}}$, where $\{X_i,X_j\}$ can be found in Table \ref{conalg}.  These operators are conformal Killing vectors of the Laplace-Beltrami operator $L_b=\psi(x,y) (\pa_x^2+\pa_y^2)$, corresponding to the general $H$ of (\ref{2d-gen}), satisfying $[L_i,L_b]=\lambda_i L_b$, where $\lambda_i$ are given by (\ref{lambdai}).

In coordinates $(u,v)$, we have
\bea
&&  L_1=e^{-v} (\cos u \pa_u + \sin u \pa_v),\quad L_2 = e^{-v} (\cos u \pa_v - \sin u \pa_u),\quad L_3 = \pa_u, \nn\\[-1mm]
&&   \label{con-algL-uv}  \\[-1mm]
&&  L_4 = \pa_v, \quad L_5 = e^{v} (\sin u \pa_v-\cos u \pa_u), \quad L_6 = e^{v} (\sin u \pa_u + \cos u \pa_v).   \nn
\eea
\end{subequations}

\subsection{The Quantum Darboux-Koenigs Cases}

We now carry out this procedure for each of the Darboux-Koenigs cases of Sections \ref{X2H=0} to \ref{X4g246}.  The only change to the formulae is the symmetrisation of products, together with the addition of a few lower order terms to some formulae.

\subsubsection{The Quantum Darboux-Koenigs Case $D_1$}

We just take the formulae \underline{directly} from Section \ref{X2g123}, giving $L_b=\frac{1}{\alpha x+\beta} (\pa_x^2+\pa_y^2)$.  However, since $L_i$ are operators (not just functions on phase space), we replace a simple product by a {\em symmetric product}, although this is unnecessary when two operators commute.  We have Killing vector $L_2$ and define
\bea
F_1 &=& L_1 L_2 -\frac{\alpha}{2} y L_b = \pa_x\pa_y-\frac{\alpha y}{2 (\alpha x+\beta)} \, (\pa_x^2+\pa_y^2),\nn\\
 F_2 &=& \frac{1}{2}(L_2L_3+L_3L_2)-\frac{\alpha}{4} y^2 L_b = y\pa_x\pa_y-x \pa_y^2+\frac{1}{2}\pa_x-\frac{\alpha y^2}{4 (\alpha x+\beta)} \, (\pa_x^2+\pa_y^2), \nn
\eea
which satisfy
$$
  [L_2,F_1]= -\alpha L_b,\quad [L_2,F_2]=  F_1,\quad [F_1,F_2]=  -2 L_2^3+\beta L_b L_2,\quad F_1^2 +\alpha F_2 L_b+ L_2^4 -\beta L_b L_2^2 = 0.
$$

\subsubsection{The Quantum Darboux-Koenigs Case $D_2$}

We now take the formulae from Section \ref{X2g246}, with Killing vector $L_2$, giving:
\bea
L_b &=& \frac{x^2}{a_2-a_1 x^2} (\pa_x^2+\pa_y^2), \nn\\
F_1 &=& \frac{1}{2} (L_2 L_4+L_4L_2) +a_1 y L_b = x\pa_x\pa_y+y \pa_y^2+\frac{1}{2}\pa_y + \frac{a_1x^2 y}{a_2-a_1 x^2} \, (\pa_x^2+\pa_y^2),\nn\\
 F_2 &=& \frac{1}{2}(L_2L_6+L_6L_2)  + a_1 y^2 L_b = 2 x y\pa_x\pa_y+(y^2-x^2) \pa_y^2+x\pa_x+y\pa_y + \frac{a_1x^2 y^2}{a_2-a_1 x^2} \, (\pa_x^2+\pa_y^2), \nn
\eea
which satisfy
\bea
&&  [L_2,F_1]= a_1 L_b+L_2^2,\quad [L_2,F_2]= 2 F_1,\quad [F_1,F_2]=  L_2F_2+F_2L_2+2 a_2 L_2L_b-\frac{1}{2} L_2,\nn\\
&&     F_1^2 = a_1 F_2 L_b+a_2 L_b L_2^2+\frac{1}{3} (L_2^2F_2+L_2F_2L_2+F_2L_2^2)+\frac{a_1}{6} L_b + \frac{11}{12} L_2^2.  \nn
\eea
We see that the {\em leading order terms} in each equation are identical to those of the classical case, but that some {\em lower order} terms were added in the last two formulae.

\subsubsection{The Quantum Darboux-Koenigs Case $D_3$}

We now take the formulae from Section \ref{X3g123}, with Killing vector $L_3$, giving:   {\small
\bea
L_b &=& \frac{e^{-v}}{c_1 e^v-2\alpha}\, (\pa_u^2+\pa_v^2), \nn\\
F_1 &=& \frac{1}{2} (L_1 L_3+L_3L_1) + \alpha \cos u\, L_b = \frac{1}{2}e^{-v} \left(\cos u\, (2\pa_u^2+\pa_v)+\sin u\,(2 \pa_u\pa_v-\pa_u) + \frac{2\alpha\cos u}{c_1 e^v-2\alpha}\, (\pa_u^2+\pa_v^2) \right),\nn\\
 F_2 &=& \frac{1}{2} (L_2 L_3+L_3L_2) - \alpha \sin u\, L_b = \frac{1}{2}e^{-v} \left(\cos u\, (2 \pa_u\pa_v-\pa_u)-\sin u\,(2\pa_u^2+\pa_v) - \frac{2\alpha\sin u}{c_1 e^v-2\alpha}\, (\pa_u^2+\pa_v^2) \right), \nn
\eea   }
which satisfy
$$
[L_3,F_1]= F_2,\quad [L_2,F_2]= - F_1,\quad [F_1,F_2]=  -c_1 L_3L_b,  \quad  F_1^2+F_2^2 = c_1 L_3^2\, L_b+ \alpha^2 L_b^2+\frac{c_1}{4} \, L_b .
$$
Again, we see that the {\em leading order terms} in each equation are identical to those of the classical case, but a {\em lower order} term is added to the last formula.

\subsubsection{The Quantum Darboux-Koenigs Case $D_4$}

We now take the formulae from Section \ref{X4g246}, with Killing vector $L_4$, giving:   {\small
\bea
L_b &=& \frac{\sin^2 u}{c_1 -2\alpha \cos u}\, (\pa_u^2+\pa_v^2), \nn\\
F_1 &=& \frac{1}{2} (L_2 L_4+L_4L_2) - \alpha e^{-v}\, L_b = \frac{1}{2}e^{-v} \left(\cos u\, (2\pa_v^2-\pa_v)-\sin u\,(2 \pa_u\pa_v-\pa_u) - \frac{2\alpha\sin^2 u}{c_1 -2\alpha \cos u}\, (\pa_u^2+\pa_v^2) \right),\nn\\
 F_2 &=& \frac{1}{2} (L_4 L_6+L_6L_4) - \alpha e^v\, L_b = \frac{1}{2}e^{v} \left(\cos u\, (2\pa_v^2+\pa_v)+\sin u\,(2 \pa_u\pa_v+\pa_u) - \frac{2\alpha\sin^2 u}{c_1 -2\alpha \cos u}\, (\pa_u^2+\pa_v^2) \right), \nn
\eea   }
which satisfy
\bea
&&  [L_4,F_1]= -F_1,\quad [L_4,F_2]= F_2,\quad [F_1,F_2]=  -2c_1 L_4L_b+4 L_4^3+\frac{1}{2} L_4,  \nn\\
&&    F_1 F_2+F_2 F_1 = 2\alpha L_b^2+ 2 L_4^4-2 c_1 L_4^2 L_b-\frac{c_1}{2} \, L_b+\frac{5}{2} L_4^2.  \nn
\eea
Again, we see that the {\em leading order terms} in each equation are identical to those of the classical case, but {\em lower order} terms are added to the last two formulae.

\subsection{The Quantum Case with Third Order Integrals}

Here we consider the quantum case of the algebra discussed in Section \ref{cubic-int}.  The strategy will again be to take the formulae {\em directly} from the classical case.  However, this time we find that we are forced to a constant curvature case, so presumably it is necessary to look at a more general conformal invariant.  Given that we are reduced to the constant curvature case, we can isolate a 3 dimensional subalgebra of {Killing vectors} from the 6 dimensional conformal algebra and express $F_1$ and $F_2$ in terms of these.

\subsubsection{The Operator Algebra}

We start with $L_b$ and the operator version of (\ref{F1-1}):
\begin{subequations}
\bea
L_b &=& \varphi(x) (\pa_x^2+\pa_y^2),    \label{Lb3}  \\
F_1 &=& \frac{c_0}{2} (L_2(y L_b)+y L_b L_2) +\frac{c_1}{2} (L_4 L_2^2+L_2^2 L_4) \nn \\
&&  \hspace{3cm} + c_2 L_1 L_2^2 +\frac{1}{2} (b(x) L_b L_1+L_1(b(x) L_b)),   \label{F13}
\eea
\end{subequations}
where the first and last symmetric sum is required as a consequence of $[L_2,y L_b]\neq 0$ and $[L_1,b(x) L_b]\neq 0$. We then have $[L_2,F_1]=c_0L_b L_2+c_1 L_2^3$, and the condition $[L_b,F_1]=0$ implies
$$
\varphi = b^2, \qquad b_x = \frac{b(c_1+c_0 b^2)}{c_2+c_1 x+b^3} \qquad\mbox{\underline{and}}\qquad b_{xxx}=0.
$$
The two conditions on $b(x)$ imply
$$
3 b^2 (c_1 b-c_0(c_2+c_1 x))(c_1+c_0 b^2) (4 c_0 b^5+8 c_1 b^3-5 c_0(c_2+c_1 x) b^2-c_1(c_2+c_1 x)) = 0.
$$
Since the scalar curvature of the corresponding metric is given by $R=2 (b b_{xx}-b_x^2)$, we must choose the quintic factor to be zero, in order to avoid {\em constant curvature}.  However, since $b_{xxx}=0$ gives $b(x) = \alpha_0+\alpha_1 x+\alpha_2 x^2$, this quintic quickly leads to $\alpha_2=\alpha_1=\alpha_0=0$.  We are thus led to choosing
\be\label{linear-b}
b(x) = \frac{c_0}{c_1}(c_2+c_1 x) \quad\Rightarrow\quad R = -2 c_0^2.
\ee
The quantum version of formula (\ref{F2-1}) for $F_2$ is  {\small
\be\label{F23}
F_2 = \frac{c_0}{4} (L_2(y L_b)+y L_b L_2)+\frac{c_1}{4} (L_6 L_2^2+L_2^2L_6)+\frac{c_2}{2}(L_3 L_2^2+L_2^2 L_3) +\frac{1}{2} (b L_b L_3+L_3(b L_b))+b_1(x) L_b L_2.
\ee  }
We have $[L_2,F_2]=F_1$ and requiring $[L_b,F_2]=0$ implies
$$
b_{1x} = x b_x = c_0 x \quad\Rightarrow\quad b_1(x) = \frac{1}{2} c_0 x^2.
$$
The final bracket is
\be\label{f1f2-q}
[F_1,F_2] = c_0 L_b F_2+\frac{3}{2} c_1 (L_2^2 F_2+F_2 L_2^2) + d_1 L_b^2 L_2+d_2 L_b L_2^3+d_3 L_2^5-\frac{3}{2} c_0c_1 L_b L_2-\frac{7}{2} c_1^2 L_2^3,
\ee
where
$$
d_1 = \frac{2 c_1^3-c_0^3c_2^2}{c_0 c_1^2},\quad d_2 = \frac{2 c_1^3-4c_0^3c_2^2}{c_0^2 c_1}, \quad d_3 = -3 c_2^2.
$$
Comparing this with (\ref{f1f2-cub}), we see that $d_1$ and $d_2$ are now determined and there are some lower order ``corrections''.  The constraint (\ref{con-cub}) now takes the form
$$
F_1^2-c_0(L_2F_2+F_2L_2)L_b-c_1(L_2^3F_2+F_2L_2^3)= L_b^3-c_2^2 L_2^6+d_1 L_b^2L_2^2+\frac{1}{2}d_2 L_bL_2^4-3 c_0 c_1 L_b L_2^2-4 c_1^2 L_2^4,
$$
where $d_1,\, d_2$ have the values given above.

\subsubsection{Additional Killing Vectors}

Given that the metric has constant curvature, it must have a 3 dimensional isometry group.  We can determine the Killing vectors quite simply by considering
$$
[K,L_b]=0, \qquad\mbox{with}\qquad  K = \sum_{i=1}^6 \alpha_i L_i,
$$
where $L_i$ are just the conformal vectors (\ref{con-algL-xy}).  We find
\bea
&&  K_1 = L_2=\pa_y, \qquad K_2 = c_2 L_1+c_1 L_4 = (c_2+c_1 x) \pa_x+c_1 y \pa_y,\nn\\[-1mm]
  &&   \label{Ki}    \\[-1mm]
&&  K_3 = 2 c_2 L_3+c_1 L_6 = 2 (c_2+c_1 x) y \pa_x +(c_1(y^2-x^2)-2 c_2 x), \pa_y\nn
\eea
which satisfy
$$
[K_1,K_2]=c_1 K_1,\quad [K_1,K_3] = 2 K_2, \quad  [K_2,K_3]=c_1 K_3-2 c_2^2 K_1,
$$
with Casimir
$$
{\cal C}_K = c_1 (K_1K_3+K_3K_1)-2 K_2^2-2 c_2^2 K_1^2 = -\, \frac{c_1^2}{c_0^2}\, L_b.
$$
In terms of these we have  {\small
\bea
F_1 &=& \frac{c_0^3}{c_1^3}\, K_2^3-\frac{c_0^3}{6 c_1^2}\,(K_1K_2K_3+K_2K_3K_1+K_3K_1K_2+K_2K_1K_3+K_1K_3K_2+K_3K_2K_1)  \nn\\
&&  \hspace{8cm} +\frac{c_1^3+c_0^3c_2^2}{2 c_1^3} (K_1^2K_2+K_2K_1^2)+\frac{c_0^3}{3 c_1}\, K_2,\nn\\
F_2 &=& \frac{c_0^3}{4c_1^3}\,(K_2^2K_3+K_3K_2^2)-\frac{c_0^3}{4 c_1^2}\,(K_1K_3^2+K_3^2K_1)+\frac{c_1^3+c_0^3c_2^2}{4 c_1^3}\, (K_1^2K_3+K_3K_1^2)+\frac{c_0^3}{4 c_1}\,K_3-\frac{c_0^3c_2^2}{2 c_1^2}\, K_1.\nn
\eea  }

\br
It is not strictly necessary to symmetrise these formulae, but not doing so does add more lower order terms.
\er

\section{Conclusions}

For metrics of constant curvature (including flat), all (polynomial in momenta) first integrals of the geodesic equations can be written as polynomial functions of the Noether constants of the isometry algebra.

In this paper, we have considered a similar procedure for building integrals in the {\em conformally flat} case from polynomial functions of {\em conformal symmetries}.  We considered the standard 6 dimensional algebra of conformal Killing vectors, together with additional quadratic elements of the form $\sigma(x,y) H$.  This approach allowed us to ``derive'', in a simple way, the 4 known Darboux-Koenigs metrics (with 1 linear and 2 quadratic integrals), as well as one of the cases derived in the classification of \cite{11-3} (with 1 linear and 2 cubic integrals).

This construction gives us a better understanding of the origin of these quadratic and cubic integrals in the classical case.  However, it also gives us a mechanism for building quantum analogues, with very little additional calculation.

In this paper, we only considered the 2 dimensional case, since our purpose was to understand the known results in this case.  However, the method should be easily extended to higher dimensional metrics, although the calculations will be considerably more complicated.  In 3 dimensions the constant curvature metrics have 6 dimensional isometry algebras and there exist conformally flat metrics with 3 dimensional isometry algebras, out of which it is possible to construct 3 quadratic invariants.  Such an example was given in \cite{f18-1}, with a closed Poisson algebra with only 4 independent functions (including the Hamiltonian).  No further (independent) integrals can be constructed from the isometry algebra, but perhaps the methods of this paper could give us the fifth integral needed for {\em maximal super-integrability}.

This paper was only concerned with integrals of the geodesic equations (ie building {\em Killing tensors}), and not with finding potential functions, consistent with integrability.  This problem was completely solved in the case of Darboux-Koenigs metrics in the papers \cite{02-6,03-11}.


\end{document}